\documentclass{aa}
\usepackage[varg]{txfonts}
\usepackage{graphicx}
\usepackage{natbib}
\bibpunct{(}{)}{;}{a}{}{,}

\usepackage[colorlinks=true,urlcolor=blue,citecolor=blue,linkcolor=blue]{hyperref}

\def\kms {{\mathrm{km}\,\mathrm{s}^{-1}}}

\begin{document}

\title{RH 1.5D: a massively parallel code for multi-level radiative transfer with partial frequency redistribution and Zeeman polarisation}
\titlerunning{{RH 1.5D: a massively parallel radiative transfer code}}

\author{Tiago M. D. Pereira\inst{1, 2, 3} \and Han Uitenbroek\inst{4}}
\institute{Institute of Theoretical Astrophysics, University of Oslo, P.O. Box 1029 Blindern, N--0315 Oslo, Norway\\ email: \texttt{tiago.pereira@astro.uio.no}
\and NASA Ames Research Center, Moffett Field, CA 94035, USA
\and Lockheed Martin Solar and Astrophysics Laboratory, Lockheed Martin Advanced Technology Center, Org. A021S, Bldg. 252, 3251 Hanover St., Palo Alto, CA 94304, USA
\and National Solar Observatory, Sacramento Peak, P.O. Box 62, Sunspot, NM 88349, USA} 

\date{Received 9 August 2014 / Accepted 4 November 2014}

\abstract{The emergence of three-dimensional magneto-hydrodynamic (MHD) simulations of stellar atmospheres has sparked a need for efficient radiative transfer codes to calculate detailed synthetic spectra. We present RH 1.5D, a massively parallel code based on the RH code and capable of performing Zeeman polarised multi-level non-local thermodynamical equilibrium (NLTE) calculations with partial frequency redistribution for an arbitrary amount of chemical species. The code calculates spectra from 3D, 2D or 1D atmospheric models on a column-by-column basis (or 1.5D). While the 1.5D approximation breaks down in the cores of very strong lines in an inhomogeneous environment, it is nevertheless suitable for a large range of scenarios and allows for faster convergence with finer control over the iteration of each simulation column. The code scales well to at least tens of thousands of CPU cores, and is publicly available. In the present work we briefly describe its inner workings, strategies for convergence optimisation, its parallelism, and some possible applications.}


\maketitle

\section{Introduction}

The field of stellar atmospheres has undergone a dramatic change with the advent of three-dimensional, radiative magnetohydrodynamic models. Complex simulations have been developed to study a variety of topics, including but not limited to solar convection \citep{SteinNordlund1998, Wedemeyer:2004, Vogler:2005}, stellar and solar abundances \citep{Asplund:1999, Asplund2000b,Allende-Prieto:2002, Collet:2007, Caffau:2007aa}, solar surface magnetism  \citep{Stein:2006aa, Cheung:2007aa, Martinez-Sykora:2008aa, Rempel:2009}, solar chromospheric dynamics \citep{Hansteen:2007aa, Martinez-Sykora:2009, Carlsson:2010aa}, convection and granulation across the HR diagram \citep{Ludwig:2006aa, Magic:2013aa, Trampedach:2014aa}. While some of these simulations include a detailed treatment of radiation, it is still not detailed enough (and would be prohibitively expensive) for the calculation of spectral lines in suitable detail. Therefore, the need arises to calculate the predicted spectra from such model atmospheres. To that effect, a variety of codes have been developed \citep[e.g.][]{Ludwig:2008aa, Leenaarts:2009a, Hayek:2011, Stepan:2013} to calculate synthetic spectra from 3D simulations, each code having their strengths and weaknesses \citep[see the review by][]{Carlsson:2008}. These codes operate under the usually valid assumption that the detailed spectral calculations will not affect the radiative magneto-hydrodynamics (MHD) calculations significantly, and one can just use the output of the simulations for the detailed spectral calculations. In the present work we describe yet another code to calculate spectra from 3D models. We believe it to be sufficiently general and particularly unique to appeal to a large community. 

RH 1.5D is derived from the RH code \citep{Uitenbroek:2001} and shares a very large code base with it. However, there are important improvements both in the optimisation of convergence and in the parallelism that merit a separate description of the code. While RH can be used to solve problems in a variety of geometries (1D, 2D, 3D, spherical), RH 1.5D is designed to solve a specific class of problems: the calculation of spectra from simulations on a column-by-column basis, or in 1.5D. The simulations can be 1D, 2D, or 3D, but the calculations are made independently for each simulation column. There are both limitations and advantages of this approach, and we discuss them below. 

The code is publicly available to download via a \emph{git} repository at \url{https://github.com/tiagopereira/rh}, with documentation at \url{http://rh15d.readthedocs.org}. The code version used throughout this paper is v1.2 \citep{rh15d-v1.2}.

This paper is organised as follows. In Section \ref{sec:radtrans} we briefly describe the radiative transfer method used and strategies to improve its convergence. In Section \ref{sec:parallel} we detail the parallelisation strategy and evaluate its efficiency. In Section \ref{sec:applications} we briefly discuss possible applications of the code, and finally conclude with a summary in Section \ref{sec:summary}.

\section{Radiative transfer}
\label{sec:radtrans}

\subsection{Formal solver}

The code solves the non-local thermodynamic equilibrium (NLTE) radiative transfer problem allowing for partial frequency redistribution (PRD). It uses the multi-level accelerated $\Lambda$-iteration method developed by \citet{Uitenbroek:2001}, which is an extension of the \citet{Rybicki:1992} formalism to allow for PRD. It employs a local approximate $\Psi$ operator and preconditioning of the rate equations, and allows for overlapping radiative transitions (i.e. proper treatment of blended spectral lines). 

The code allows for angle-independent and angle-dependent PRD as detailed in \citet{Uitenbroek:2002}. As pointed out by \citet{Uitenbroek:2002}, the effects of angle-dependent PRD are typically small in hydrostatic atmosphere models, but they can be considerable in some cases for time-dependent hydrodynamic models. Unfortunately, the calculation of the angle-dependent redistribution function comes with a significant computational cost, which is not desirable for large 3D simulations. To mitigate for this limitation \citet{Leenaarts:2012} developed a fast approximation for angle-dependent PRD in moving atmospheres, and this hybrid mode is also implemented in the code. The implementation allows for PRD to be enabled for some or all bound-bound transitions in an atom, the treatment reverting to complete redistribution (CRD) when PRD is switched off.
Besides supporting overlapping radiative transitions, the code allows for solving the rate equations of several atoms concurrently, consistently treating any overlapping transitions. Molecular rotation-vibration transitions can also be calculated, using the method of \citet{Uitenbroek:2000}. In addition to the transitions and continua calculated in NLTE, the code allows for the inclusion of additional spectral lines calculated assuming LTE. These additional lines are added as extra background opacity and can therefore overlap with any existing transition treated in NLTE. 

The code can also include the effects of Zeeman splitting in bound-bound atomic or molecular transitions, allowing for the calculation of the full Stokes vector of polarised radiation.

\subsection{Column-by-column approach}
\label{sec:1.5d}

By solving the radiative transfer problem on a column-by-column basis (1.5D) one is neglecting the effect of inclined rays in a 2D or 3D problem. This has two disadvantages. First, one can only calculate the emergent spectra in the vertical direction from a given simulation. Second, neglecting the inclined rays in the calculation of the angle-averaged mean intensities will result in a different mean radiation field (higher in hotter locations, lower in cooler regions), affecting the source function and therefore the level populations and resulting intensity. This effect will depend crucially on the photon mean free path. When the mean free path is smaller (e.g. denser regions where the optical depth reaches unity in the continuum or weak lines) the inclined rays will not progress much farther than one simulation grid interval and their effect is negligible. Conversely, in the less dense regions where the cores of strong lines are formed, the mean free path can be large enough that the inclined rays can on average travel sideways through several simulation cells and influence the mean radiation field. The suitability of the 1.5D approximation will therefore depend on the problem to be studied and the required accuracy. Several examples in the literature indicate that the differences between 1.5D and full 3D NLTE radiative transfer in stellar atmospheres are generally small and occur mostly in the cores of very strong lines \citep{Kiselman:1995, Leenaarts:2010, Leenaarts:Mg1, Holzreuter:2013}.

Taking into account its limitations, there are several advantages in using a 1.5D approach. Because each column can be treated individually, one can tailor the iteration and convergence optimisation strategies to each column in a simulation. Doing this can lead to faster calculations, and the iterations are no longer limited by the slowest converging points in the simulations. It also allows one to deal with problematic regions more effectively, spending more computational power where it is needed the most. In full 3D calculations a single problematic column can slow down the whole computation, or in extreme cases lead to non-convergence of the global solution. This problem is particularly relevant to PRD calculations in simulations of the solar chromosphere. At the time of writing, PRD calculations in chromospheric lines are still not possible in full 3D due to numerical instabilities induced by the strong temperature and velocity gradients present in such models. A non-convergent column in the 1.5D approach can be dealt with methodically and independently of the rest of the columns. The 1.5D problem also parallelises very well, given the large number of columns in typical simulations.

\subsection{Convergence optimisation}

\begin{figure}
\begin{center}
\includegraphics{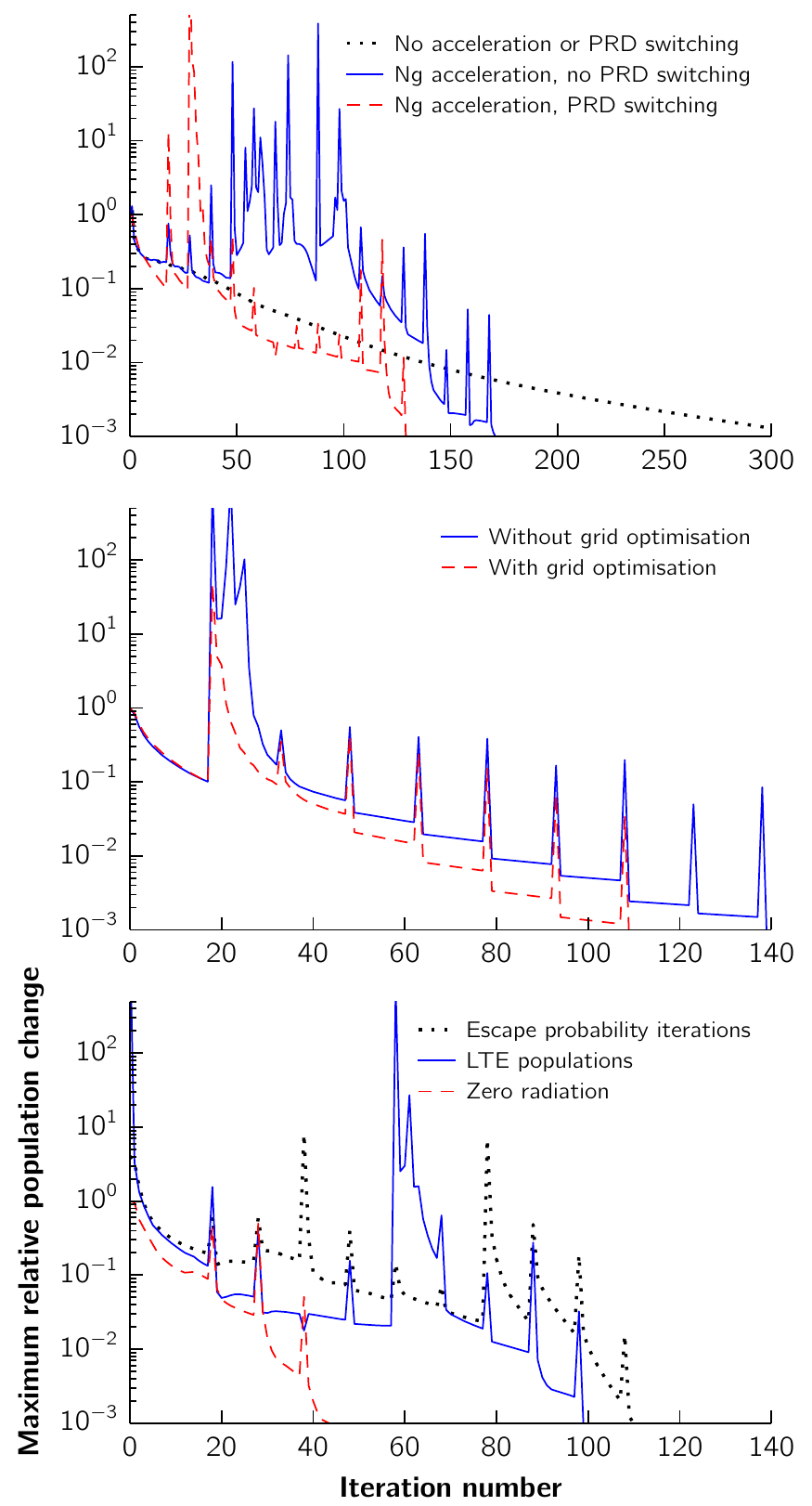}
\end{center}
\caption{Effect of different optimisation options in the convergence behaviour. Each panel was calculated using a 5-level Mg~\textsc{ii} atom with PRD in the h\&k lines, for different columns of a 3D MHD simulation. \emph{Top panel:} effect of Ng acceleration and PRD switching. \emph{Middle panel:} effect of grid optimisation. \emph{Bottom panel:} effect of different starting solutions.\label{fig:convergence}}
\end{figure}

The formal solver used has good convergence properties, as tested by \citet{Uitenbroek:2001}. To accelerate the convergence it can use the method of \citet{Ng:1974} with a general order \citep{Auer:1987aa, Auer:1994aa}, both for the $\Lambda$-iterations and the PRD iterations. To further improve the convergence speed and ensure all columns converge, there are several additional methods available in the code that we detail below. 

\paragraph{Different starting solutions} To a large extent, the time any iterative solver takes to achieve convergence will depend on how far the starting solution is from the end result. To ensure this proximity, the code allows for three different methods to estimate the initial level populations: LTE populations, zero radiation, and by doing escape probability iterations. With zero radiation one assumes a mean radiation field of zero and then solves the statistical equilibrium equations for the level populations. This approach works better for atoms with strong lines, whose level populations have significant departures from LTE. With escape probability iterations one assumes a certain probability of photons escaping and obtains the populations by iteratively solving the statistical equilibrium equation for a few iterations \citep[see][and references therein]{Hubeny:2001aa}.

\paragraph{Grid optimisation} The original depth grid from a simulation can be optimised for the radiative transfer calculations, where strong temperature and velocity gradients can cause numerical instabilities. The code allows for interpolation of the original atmospheric variables into an optimised grid where more points are placed in regions of increased temperature, velocity, or optical depth gradients. It is also possible to exclude the higher parts of a simulated atmosphere when these are not relevant for the calculations (e.g. not including the transition region or coronal parts when calculating lines formed in the photosphere). This minimises any numerical instabilities that might occur when strong gradients occur in such regions. The grid optimisation for each column is another advantage of the 1.5D approach: 3D radiative transfer codes typically assume a Cartesian mesh with the same fixed depth scale for all columns. 

\paragraph{Collisional-radiative switching} The code can make use of the collisional-radiative switching method of \citet{Hummer:1988}, which can improve convergence by including NLTE effects gradually. By definition, under LTE conditions collisional excitations dominate over radiative excitations. With collisional-radiative switching one artificially increases the atomic collisional rates by an arbitrary multiplication factor so that they are comparable to the radiative rates and the solution is close to LTE. A good multiplication factor will depend on the atom and atmosphere used, and can usually be between $10$ to $10^{5}$. As the iterations progresses, this factor is decreased until it reaches unity and full NLTE is achieved. For some problems this approach greatly increases the convergence speed.

\paragraph{PRD switching} Similarly to collisional-radiative switching one can also introduce the effects of PRD gradually, to avoid strong non-linearities and improve convergence. Following an idea from J. Leenaarts \& H. Uitenbroek (2010, priv. comm.), this was implemented in the code by modifying the ratio $\rho(\nu, \mathbf{n}) = \psi(\nu, \mathbf{n}) / \phi(\nu, \mathbf{n})$, where $\psi(\nu, \mathbf{n})$ and $\phi(\nu, \mathbf{n})$ are the PRD emission and CRD absorption/emission line profiles, respectively \citep[see Eq. 14 of][]{Uitenbroek:2001}. For each transition, frequency and direction, $\rho$ is scaled in the following way:
\begin{equation}
  \rho' = \rho \epsilon + (1 - \epsilon).
\end{equation}
When $\epsilon=0$, $\psi(\nu, \mathbf{n}) = \phi(\nu, \mathbf{n})$ and one has CRD.
We initialise $\epsilon$ with a small value $\epsilon_0$ and at each iteration multiply it by $n_{i}^{2}$, the squared number of the current iteration, until it reaches unity. Therefore, $\epsilon \to 1$ and $\rho' \to \rho$ in $\approx\epsilon_0^{-1/2}$ iterations.  

In Figure~\ref{fig:convergence} we show some examples of the effects of different convergence optimisation options on the convergence speed, given by the number of iterations to reach a certain threshold in the maximum relative population change. Each panel was taken from different columns of a 3D MHD simulation \citep[the same used by][]{Pereira:Mg3}, for a 5-level Mg~\textsc{ii} atom with PRD in the h\&k lines. The top panel was calculated from a column particularly difficult to converge -- with no acceleration it takes more than 300 iterations to reach the $10^{-3}$ level, which is improved by the use of Ng acceleration and further even by using PRD switching. The middle panel shows the effects of interpolating the height grid to one better optimised for radiative transfer. This grid optimisation yields an improved convergence in the majority of cases. Finally, the bottom panel highlights how important it is to choose an appropriate starting solution. In this case, the zero radiation solution was considerably better, but this varies with the atom and atmosphere model used.

Taking full advantage of the 1.5D approach one can use different options for each column, or run first with an aggressive choice of options to ensure a fast convergence of a great number of columns, and then re-run with different options for the columns that did not converge.

\section{Parallelism}
\label{sec:parallel}

\subsection{Parallelisation strategy}

Given that each column is a completely independent calculation, the problem lends itself very well to parallel computing. A very limited amount of communication is necessary between different processes (typically CPU cores). The code has been parallelised using the message-passing interface library (MPI\footnote{\url{http://www.mpi-forum.org/docs/mpi-3.0}}), and has two running modes: \emph{normal} and \emph{pool}. 

In the \emph{normal} mode the total number of columns to be calculated (tasks) is divided by the number of processes, so the amount of tasks each process has is about the same and is known in the beginning of the execution. Each process starts working through its task list until it is finished, and then writes the output to the disk and waits for the other processes to complete. The advantage of the \emph{normal} mode is that it is conceptually simpler (each process executes essentially the same code) and the input/output overhead is smaller because the results are written only once, at the end. A disadvantage of the \emph{normal} mode is that while the number of tasks is approximately the same between processes, not all tasks take the same amount of time to complete. This means that the whole execution will have to wait on the slowest process, the process whose task list included columns that took longer to converge. In some problems these slower processes can take more than twice of the typical running time. Because the output is only written at the end, each process needs to save the intermediate results in memory before they are written to disk. This has the disadvantage of additional memory requirements: if processes have a very large number of tasks to complete, this extra memory can be substantial or even prohibitively expensive.

In the \emph{pool} mode there is one \emph{manager} process and many \emph{worker} processes. The role of the \emph{manager} process is to distribute a \emph{pool} of tasks to be calculated by the other processes. A \emph{worker} process solves the problem for a given simulation column, writes the output do disk, and asks the \emph{manager} for more work, until all tasks are completed. This mode amounts to dynamic load balancing, where a process that has to calculate a column that is slower to converge will not slow down the overall calculation because it will not get more work until it is done. Therefore the load distribution is optimal with the \emph{pool} model. The disadvantage of the \emph{pool} mode is that its input/output patterns are more intensive, because the output is written immediately after each column is finished. On the other hand, the memory footprint is also minimised because unlike in the \emph{normal} mode, no saving of intermediate results is necessary. The \emph{pool} mode also requires one extra process for the \emph{manager}. 

\subsection{Input/Output strategy}

While the computational power of modern systems has undergone a huge increase following Moore's law \citep[see][]{mooreslaw}, the speed of hard drives has not increased by as many orders of magnitude, instead increasing slowly\footnote{At the time of writing, solid-state device (SSD) hard drives are not widespread for storing large amounts of data.}. This means that increasingly the input/output (I/O) performance of computer codes gets more important, as I/O could often be the bottleneck of an otherwise well performing code. I/O performance is particularly important in 3D NLTE radiative transfer codes, because the output can be very large. Radiative transfer variables such as the optical depth that are a function of space and wavelength can take up a lot of memory for large simulations. Assuming a typical simulation with $1024^3$ grid points, and a calculation with 500 frequency points, storing a single of these variables will take 2~Tb, and therefore the full output of such calculations can easily run into the tens of terabytes -- a very large amount by most standards.

The I/O strategy of RH 1.5D is a tradeoff between efficiency and memory usage. Good I/O practices involve read/writes in large blocks and if possible in a parallel collective operation with many processes. Given the very large size of the radiative transfer variables, it is not always practical to \emph{buffer} large amounts of data to write in large blocks, as the memory requirements can be very large. In RH 1.5D all processes read/write in an independent and asynchronous manner. There are but a few output files and they are written by all processes concurrently. Under the \emph{normal} mode the output writing phase takes place after a process has finished all its tasks, while under the \emph{pool} mode this occurs after each column was finished. The output files typically reside on a parallel filesystem such as Lustre\footnote{\url{http://lustre.opensfs.org/}}, where clusters with many nodes are combined to provide aggregate I/O throughput. On a typical run of RH 1.5D the concurrent writes are but a small fraction of the total number of processes. However, if the computing time is short between the write operations, many processes will be competing for the I/O resources and contention can happen (in particular if running with several thousand processes). This is an important limitation to keep in mind, and its severity will depend on the type of problem being calculated, the filesystem implementation and load per process.

Except for a few scratch files, the I/O implementation of RH 1.5D makes use of version 4 of the NetCDF library\footnote{\url{http://www.unidata.ucar.edu/netcdf/}}, which in turn is based on the HDF5 library\footnote{\url{http://www.hdfgroup.org/HDF5/}}. It fully supports parallel I/O, and can be configured to use different I/O drivers (either MPI-IO or MPI POSIX). Therefore, much of the efficiency is passed on to the MPI implementation, to which the code is agnostic. Tests have shown the adopted I/O model to be efficient and scalable, provided there is an existing high-throughput parallel filesystem in place.

\subsection{Scaling}

\begin{figure}
\begin{center}
\includegraphics{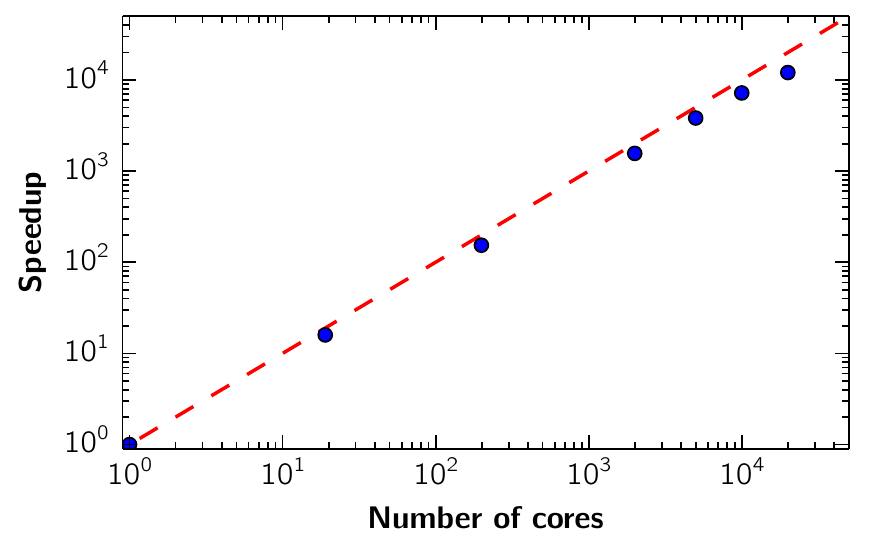}
\end{center}
\caption{Speedup of RH 1.5D, from a test calculation using a 3D MHD simulation with $504\times504$ columns (see text). Dashed line depicts linear scaling.\label{fig:scaling}}
\end{figure}

The code has been routinely run with several thousand CPU cores in different supercomputers. It has been tested with up to $20\;000$ cores and shows nearly linear scaling. In Figure~\ref{fig:scaling} we show the speedup of RH 1.5D (in pool mode) from tests ran on the Pleiades supercomputer from the NASA Advanced Supercomputing Division. The test computation involved calculating the full Stokes spectra for the 12-level Mg~\textsc{ii} atom for one snapshot of a 3D MHD simulation ran with the \emph{Bifrost} code \citep{Gudiksen:2011}. We used the same simulation and atom as \citet[][see references therein for details]{Pereira:Mg3}. The simulation has $504\times504\times467$ grid points, and the atom has 890 frequency points. The runs were carried out using the Ivy Bridge nodes of Pleiades, with Intel Xeon E5--2680v2 processors running at 2.8 GHz, each node with 2 ten-core processors and 64 Gb of RAM. The Lustre filesystem was used with 165 object storage targets (OSTs) and a stripe size of 4 Mb.

The scaling has been confirmed using other systems, such as supercomputers funded by the Research Council of Norway and smaller clusters. It is also possible to run the code in normal workstations, and the code has been tested with a variety of C compilers in several operative systems.

\section{Applications}
\label{sec:applications}

\begin{figure*}
\begin{center}
\includegraphics{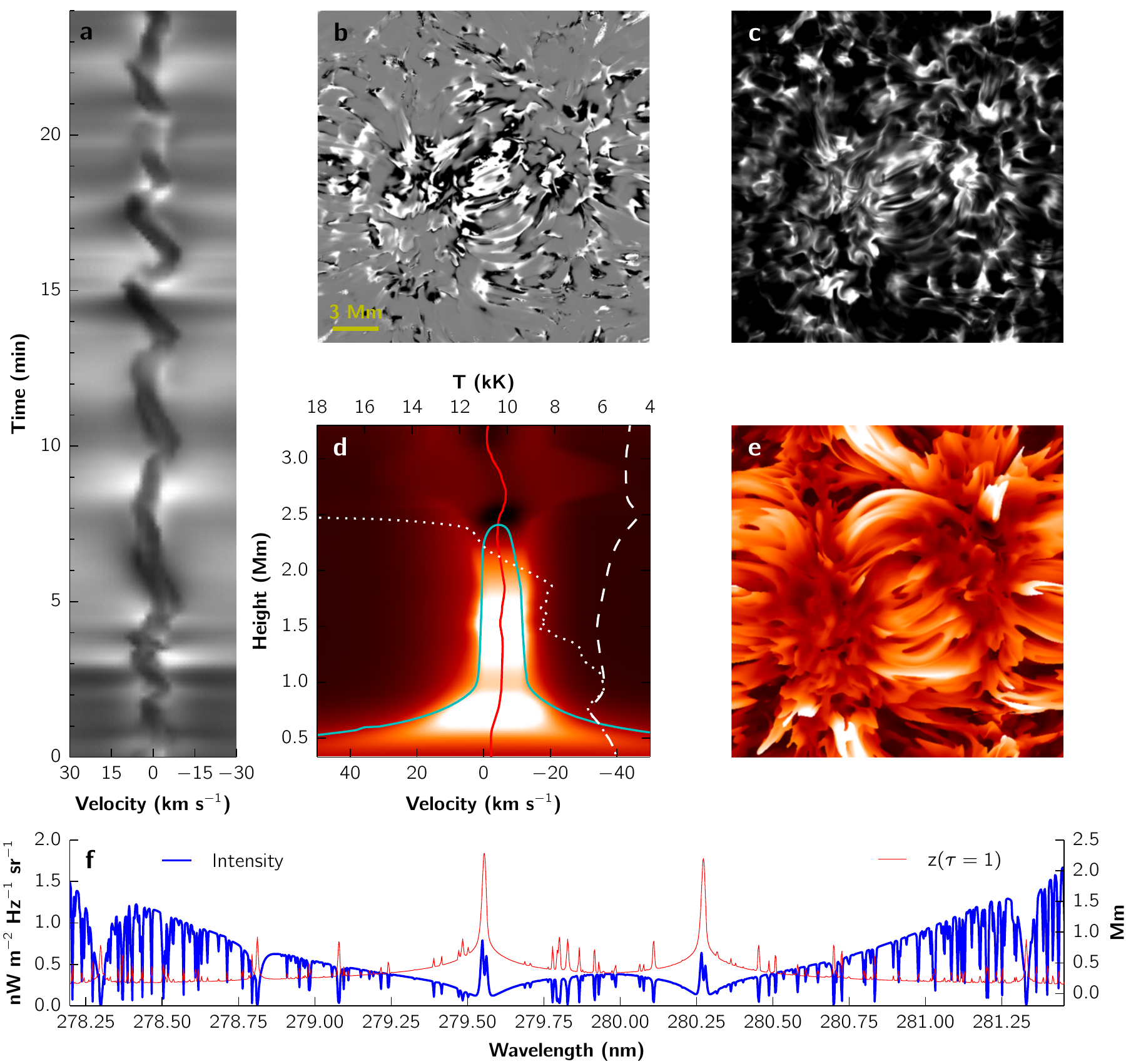}
\end{center}
\caption{Example applications of RH 1.5D, calculated from a 3D MHD simulation. Panel (a) is a velocity-time diagram from one column of the simulation, showing a shock pattern in the Ca~\textsc{ii}~H line. 
Panels (b), (c), and (e) are a top view of the $24\times24$~Mm$^2$ simulation box, showing the properties of the radiation at 279.548~nm, approximately on the mean blue peak ($k_{2V}$) feature of the Mg~\textsc{ii}~k line. 
Panel (b) shows the Stokes V intensity, a proxy for the vertical magnetic field in the chromosphere, while panel (c) shows the squared root of the radiation intensity and panel (e) shows $z(\tau=1)$, the height where the optical depth reaches unity, scaled from 0.5~Mm (\emph{black}) to 3.5~Mm (\emph{white}). Panel (d) shows the source function around the Mg~\textsc{ii}~k line for one column in the simulation, as a function of velocity and height. Additional lines are $z(\tau=1)$ (\emph{cyan}), atmospheric velocity (\emph{red}), and in brightness temperature units (in kK, top scale) the source function at $-5\;\kms$ (\emph{white dashed}), and the LTE source function (\emph{white dotted}). One can see that in this case the departures from LTE occur at around $z\approx 1$~Mm. The bottom panel (f) shows the spatially averaged spectrum around the Mg~\textsc{ii} h\&k lines from the same simulation snapshot, with many blending lines included  (\emph{thick blue}). Also included is the spatially averaged $z(\tau=1)$ as a function of wavelength (\emph{thin red}). \label{fig:app}}
\end{figure*}

RH 1.5D can be used in a broad variety of problems concerning the calculation of spectral lines from atmospheric simulations, and where 3D effects are not important (see Section~\ref{sec:1.5d}). Its strong point is the calculation of lines in NLTE and PRD, but it can just as well be used as a general NLTE code even when PRD is not necessary. The ability to include molecules in NLTE, to treat multiple atoms and molecules concurrently in NLTE, and to include an arbitrary amount of lines in LTE are other strong points that make it unique. In addition, the code can also calculate the polarisation from the Zeeman effect, important for a forward-modelling approach to understand the magnetism of stars. (Scattering polarisation calculations are not supported.). Given its very good scaling properties (see Figure~\ref{fig:scaling}), one can even use it to calculate only lines in LTE when a large or fast job is required (e.g. the calculation of the Stokes vector from very large simulations, or the calculation of spectral lines for the purpose of abundance determinations from a grid of 3D models). 

Beyond the simple calculation of intensity and its polarisation, the detailed output from the code (level populations, source functions, radiation field, etc.) can be used for detailed studies of line formation in stellar atmospheres. In Figure~\ref{fig:app} we show some example diagnostics that can be calculated from RH 1.5D, from the study of waves using Ca~\textsc{ii} lines in panel (a), to chromospheric Stokes V and Stokes I in panels (b) and (c), details of the radiative transfer in panel (d) and of the formation height in panel (e). In panel (f) we show a synthetic high-resolution spectrum around the Mg~\textsc{ii} h\&k, along with an estimate of the formation height of each wavelength. The code and such diagnostics have been recently used to study the formation properties of lines observed by the IRIS mission \citep{Leenaarts:Mg1,Leenaarts:Mg2,Pereira:Mg3}.

\section{Summary}
\label{sec:summary}

We have described RH 1.5D, a massively-parallel code for polarised multi-level radiative transfer with partial frequency distribution. It is derived from the RH code and includes important convergence optimisation features to speed up or improve convergence, which are particularly useful in dynamic models of chromospheres. While one should be aware of its limitations, the calculation of spectra using the 1.5D or column-by-column is a good approximation in many cases, and generally allows for faster convergence and more flexible methods of improving convergence. 

The code is able to scale well to several thousands of processes and, provided an efficient I/O infrastructure, there is no reason not to expect it from scaling even further. It is also publicly available and provided with a detailed documentation. 

With far-ranging applications in the field of solar and stellar spectropolarimetry (albeit only accounting for the Zeeman effect), we believe that the code will be of interest to a large part of the community, and encourage its adoption.

\begin{acknowledgements}
We are grateful to Jorrit Leenaarts for his extensive contributions to the code (in particular the PRD switching, approximate angle-dependent PRD, and support for many collisional processes). We would also like to thank Bhavna Rathore, Mats Carlsson, and Hsiao-Hsuan Lin for their contributions to the code.
We gratefully acknowledge the use of supercomputer resources provided by the NASA High-End Computing (HEC) Program through the NASA Advanced Supercomputing (NAS) Division at Ames Research Center (project s1061) and from the Notur project through grants from the Research Council of Norway.
This work was supported by the European Research Council under the European Union's Seventh Framework Programme (FP7/2007-2013) / ERC Grant  agreement No. 291058. T.M.D.P. was supported by the NASA Postdoctoral Program at Ames Research Center (grant NNH06CC03B).
\end{acknowledgements}

\bibliographystyle{aa}

\end{document}